\documentclass[9pt,twocolumn,twoside]{opticajnl}
\journal{opticajournal} 

\setboolean{shortarticle}{true}

\usepackage{lineno}

\title{Spacing and Wavelength Tunable Frequency Comb}

\author{Yijia Cai}
\author{Zhixin Liu}

\affil{Department of Electronic and Electrical Engineering, University College London, United Kingdom}
\affil{yijia.cai.19@ucl.ac.uk, zhixin.liu@ucl.ac.uk}

\begin{abstract}
Optical frequency combs (OFCs) are becoming increasingly prevalent in applications such as optical communications, radio signal processing and dual-comb spectroscopy. Many of these applications require a broad and flat spectrum with flexibility in tuning the center wavelength and the tone spacing with a maintained spectral profile. However, most OFC generators are either not tunable nor cannot maintain their spectral profiles while tuning. In this paper, we demonstrate tone spacing tunability over 25-32 GHz and wavelength tunability from 1548 to 1568 nm using a cavity-less OFC based on electro-optic comb and parametric expansion. Our comb features high power (>2~W), wideband (>90~nm), high optical signal-to-noise ratio (OSNR, >25~dB), and, importantly, a maintained spectral profile. We show that a maintained spectral profile can be achieved by primarily adjusting the power of the RF driving signal and the gain of optical amplifiers. We also quantitatively study the changes in the spectral profile. 

\end{abstract}

\setboolean{displaycopyright}{false} 

\begin{document}

\maketitle
\section{Introduction}

Originally developed to count optical cycles for frequency metrology \cite{udem2002optical}, optical frequency combs (OFC) have been increasingly explored in new applications such as optical communications\cite{ataie2015ultrahigh, Sohanpal2024ECOC}, wireless and radio communications\cite{nopchinda2023multiband}, spectroscopy\cite{picque2019frequency}, signal processing\cite{deakin2020dual}.
These new applications exploit and require features different from the classic octave-spanning bandwidth in metrology applications. For example, 
long-haul optical communications use frequency and phase-locked optical tones to mitigate nonlinear penalties more efficiently \cite{temprana2015overcoming,sohanpal2023impact}. Ultra-wideband optical communications require high power and high optical-signal-to-noise ratio (OSNR) optical tones over a bandwidth exceeding 80~nm for high-capacity wavelength division multiplexed (WDM) transmission over the telecom C and L bands \cite{puttnam20152}. 
In practical optical transmitters and wavelength routers, it is critical to have tone spacing and center wavelength tunability at GHz level, such that individual comb tones can align to the center of demultiplexer's channels, ensuring high OSNR for high capacity \cite{zhou2023communications,nopchinda2023multiband,zhou2024dual}.
Beyond optical communications, dual-frequency comb (DFC) based signal processing, such as DFC-based digitization and dual-comb spectroscopy, also requires tunable spacing with a flat and unchanged spectral profile for high-resolution digitization and high scanning rate spectrum measurement \cite{deakin2020dual,coddington2016dual}.

Most OFC generators have limited tunability in spacing and center wavelength, and such tuning is usually associated with a significant change in their bandwidth and spectral profile. Cavity-based OFC generators, such as mode-locked lasers (MLL), resonant electro-optic (EO) modulator combs, and Kerr nonlinear microring combs, have limited spacing tunability (typically within 10s of MHz), and generally produce non-flat spectral shapes due to their physical operational principles \cite{pan2020quantum,rueda2019resonant,zhang2019broadband}. Take recirculating EO combs as an example, although the tone spacing can be tuned by adjusting the RF driving frequency away from the cavity resonant frequency, it leads to a drastic change in both bandwidth and spectral shape\cite{qureshi2023tunable,zhang2023power}. 
Cavity-less OFCs, such as single-pass EO combs \cite{torres2014optical}, nonlinear parametric combs \cite{kuo2013wideband}, or a combination of the two \cite{ataie2014spectrally,cai2023design}, exhibit GHz-level spacing tunability and flat spectra. It has been shown in \cite{metcalf2015broadly, lo2022scalable} that flat spectra and continuous tuning of tone spacing over a few GHz can be achieved using modulator-based EO combs \cite{metcalf2013high}. Nevertheless, the spectral width of single-pass EO combs is limited, typically less than 20~nm, which is insufficient for ultra-wideband transmission or spectroscopy. Nonlinear parametric combs can, in principle, provide all desired features, but their spacing tunability has not been qualitatively studied. 

\begin{figure*}[h!]
\centering
\includegraphics[scale = 0.35]{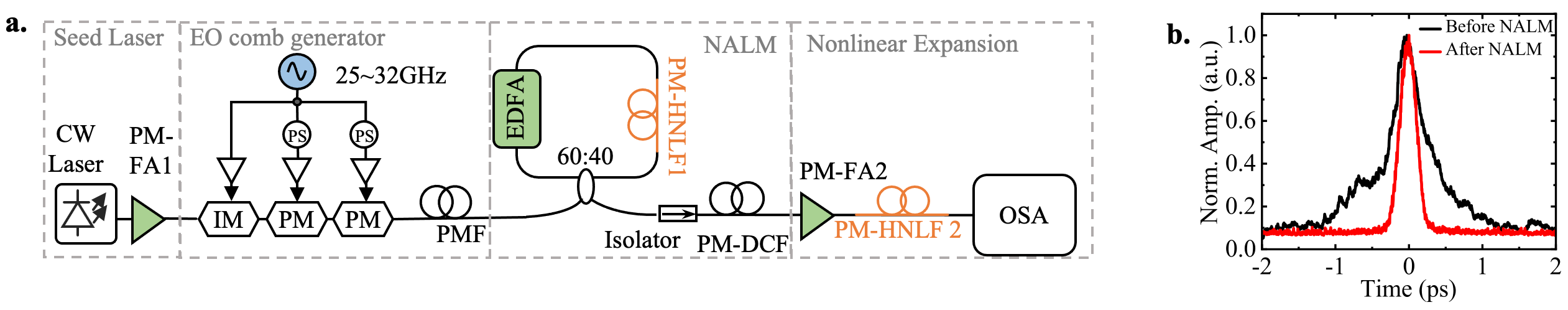}
\caption{\textbf{a.} Schematic diagram of our OFC generator. \textbf{b.} Autocorrelated trace for signal before (black) and after (red) the NALM. 
CW, continuous wave; PM, polarization maintaining; FA, fiber amplifier; IM, intensity modulator; PM, phase modulator; PS, phase shifter; NALM, nonlinear amplifying loop mirror; EDFA, erbium-doped fiber amplifier; DCF, dispersion-compensating fiber; HNLF, highly nonlinear fiber; OSA, optical spectrum analyzer; Norm. Amp., normalized amplitude.}
\label{fig:experiment setup}
\end{figure*}

In this paper, we investigate the spacing and wavelength tunability of an EO-comb pumped nonlinear parametric comb, with a focus on the quantitative study of changes in bandwidth and spectral profile. We first define the metric of spectral profile error (SPE), based on which we evaluate how well an OFC maintains its spectral profile while tuning. This evaluation is carried out at each stage to show how to optimize various parameters (RF power, RF frequency, optical power, center wavelength) to minimize the SPE. Although the design and performance of the OFC generator have been reported in our previous paper \cite{cai2023design}, the performance of spacing and wavelength tuning, as well as the optimization process for generating a maintained spectral profile, has not been studied. This study allows us to achieve continuous tone spacing tuning over a range of 25-32~GHz and center wavelength tuning of more than 20~nm, while maintaining a bandwidth of over 85~nm with a flat spectral profile, which has not been achieved in the prior art. 

\section{Experimental setup and method}
\subsection{OFC system}
The layout of our polarization-maintaining (PM) parametric comb generator is shown in Fig.\ref{fig:experiment setup}a, which consists of three stages, EO comb generator, a nonlinear amplifier loop mirror (NALM) based pulse shaper, and comb expansion with nonlinear fiber.

A CW laser source is sent through a series of modulators, each driven by the same tunable radio frequency (RF) oscillator. The IM has a $V_\pi$ and a 3-dB bandwidth of 40~GHz is biased at quadrature to carve out a flat-top pulse train. The pulse trains are modulated by cascaded phase modulators (PM) with a $V_\pi$ of 3.3~V, which generates Bessel components to yield a 10-dB bandwidth of 14~nm. RF phase shifters were used to align the phase of the RF signals that drive the IM and PM. The phase is adjusted when varying the RF frequency to compensate for the change in RF group velocity. The tone spacing is adjusted by tuning the frequency of the RF synthesizer that operates within 25-32~GHz.
The chirped pulses can be compressed to their transform limit after the same length of the compression fiber,which is a spool of 50-m standard Corning PANDA 1550-nm polarization maintained fibre (PMF). Moreover, the center wavelength of the seed laser is tuned from 1548-1568~nm to study the wavelength flexibility of the comb source.
The following pulse shaper stage suppresses the pedestals of the compressed optical pulses, producing a pulse train with a Gaussian-like pulse and spectral shape for a smooth spectrum broadening in the last stage. 
The time domain pulse shaper is a nonlinear amplified loop mirror (NALM) including an in-loop Erbium-doped fiber amplifier (EDFA) and a 40~m highly nonlinear fiber (PM-HNLF1) with a dispersion of -0.5 ps/(nm$\cdot$km), which transmits the pulse peak and reflects the pedestals of the pulses. Finally, a booster amplifier increases optical power to higher than 2 Watts power, yielding a quasi-Gaussian pulse for efficient nonlinear expansion using PM-HNLF2, which has a dispersion of -1.3 ps/(nm$\cdot$km). Both HNLF1 and HNLF2 have a similar effective gamma of about 10.5~W\textsuperscript{-1}km\textsuperscript{-1}

The pulses are characterized using an autocorrelator, assuming Gaussian pulse shape. Fig.\ref{fig:experiment setup}b shows the signal before and after the NALM. The pulse width in black is approximately 470~fs, and then pulse is compressed to 270~fs due to the self-phase modulation in the loop. The peak to pedestal ratio shows a 9~dB improvement, as indicated by the red curve. 
All spectra are measured using an optical spectrum analyzer (OSA) at a resolution of 0.02~nm. The OSNR was calculated using 0.1~nm noise bandwidth.

\subsection{Evaluation of metrics}
To quantitatively study how well the OFC generator can maintain its spectral profile while tuning, we define a metric of spectral profile error (SPE) with a reference comb spectrum. This is calculated by fitting the peak power of measured tones using a spline function and calculating the root mean square error (RMSE) of interpolated tones at the same frequencies. 
Both the aveSPE and the maxSPE (the maximum power difference of the fitted profile) are used to assess the performance, defined as Eq.\ref{eq:aveSPE} and as Eq.\ref{eq:maxSPE}
\begin{equation}\label{eq:aveSPE}
    aveSPE = RMSE = \sqrt{\frac{\sum_{i=1}^{N} (x_i-\hat{x}_i)^2}{N}}
\end{equation}
\begin{equation}\label{eq:maxSPE}
    maxSPE = max(\sqrt{(x_i-\hat{x}_i)^2})
\end{equation}
where $x_i$ and $\hat{x}_i$ are the power of the referenced tones and fitted tones, respectively. $N$ is the total number of points used in the calculation.

\section{Results and Optimization}
\subsection{EO comb}
\begin{figure}[h]
\centering
\includegraphics[scale = 0.3]{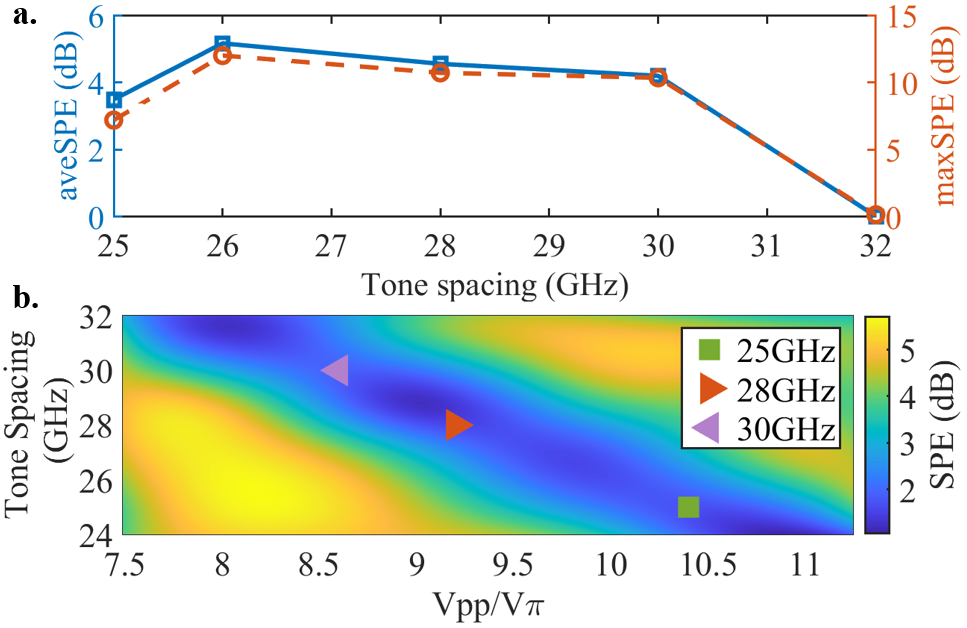}
\caption{SPE analysis for EO comb \textbf{a.} experimental aveSPE (blue, left) and maxSPE (red, right) at different tone spacings with no power optimization. \textbf{b.} simulated aveSPE (heatmap); optimum experimental aveSPE (solid markers)}
\label{fig:EOcomb_SPE}
\end{figure}

\begin{figure}[hbt!]
\centering
\includegraphics[scale = 0.31]{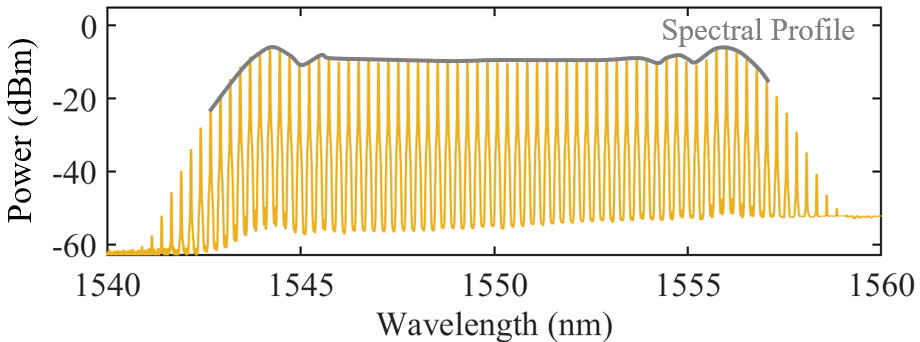}
\caption{EO comb spectrum with 32~GHz tone spacing}
\label{fig:Reference EO comb spectrum}
\end{figure}

We first analyzed the SPE of the EO comb using both experiment and simulation. As shown in \cite{metcalf2013high}, varying RF frequency with the same RF power would result in larger tone spacing and broader bandwidth. To keep the same spectral profile, we adjust the RF power driving the phase modulators. Theoretically, the comb bandwidth is determined by the number of Bessel elements, thus is determined by the ratio of the peak-to-peak driving voltage($V_{pp}$) over the $V_\pi$.
Fig.\ref{fig:EOcomb_SPE}a shows the experimental measurements of the aveSPE, obtained with the same RF power driving the phase modulators at 25-32~GHz. Using the spectrum of the 32-GHz spacing comb as the reference, we obtained an aveSPE of about 5~dB and a maxSPE of about 10~dB. Fig.\ref{fig:Reference EO comb spectrum} shows the referenced EO comb spectra 32~GHz, exhibiting 14~nm bandwidth and more than 40dB OSNR. 

By optimizing the $V_{pp}/V_\pi$, we can significantly reduce the SPE, as shown in 
Fig.\ref{fig:EOcomb_SPE}b, which shows the aveSPE obtained at different tone spacing and $V_{pp}/V_\pi$. 
By adjusting $V_{pp}/V_\pi$, we can reduce the aveSPE to less than 2~dB and maxSPE to less than 5~dB over the whole tuning region. 
In the experiment, we can obtain similar results as shown by the simulation. The markers in Fig.\ref{fig:EOcomb_SPE}b represent the experimentally measured aveSPE at tone spacings of 25, 28, and 32~GHz after optimizing the driving voltage.

\subsection{Shaped Pulses after NALM}
\begin{figure}[hbt!]
\centering
\includegraphics[scale = 0.16]{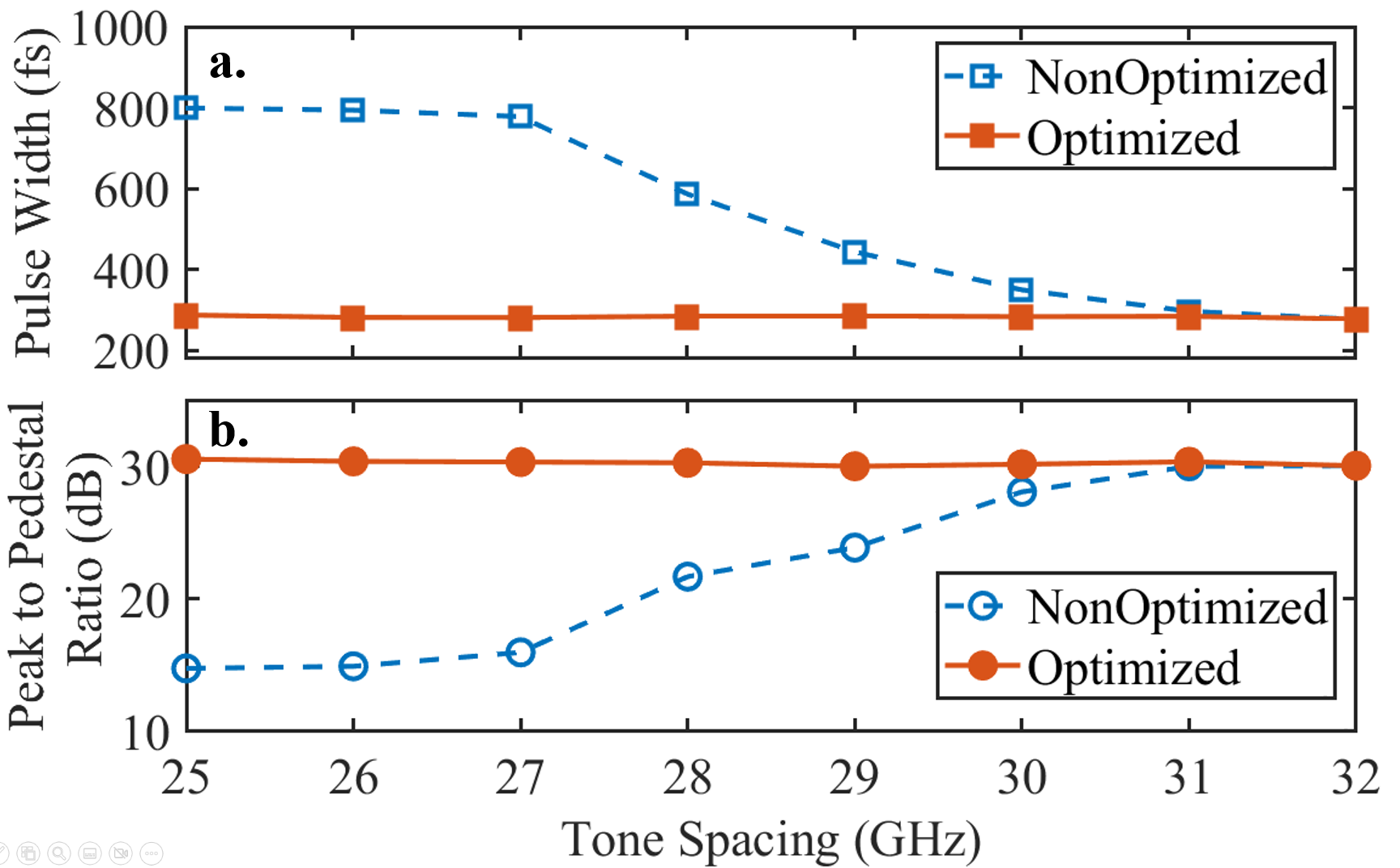}
\caption{Simulated NALM output pulses at different repetition rate.
\textbf{a.} pulse width. \textbf{b.} peak to pedestal ratio of pulses. (blue-dashed line: with no power optimization; red solid line: with power optimization)}
\label{fig:NALM Output}
\end{figure}
With a maintained EO comb spectrum, the compressed pulses after the same fiber exhibit the same pulse shapes, which have a significant pedestal, but have different pulse energy due to the varied repetition rates. As a result, maintaining the spectral profile requires preserving the same 'clean' pulse shape after the NALM. With the same NALM, different input power would result in different in-loop nonlinearity and significantly different outputs. By varying the pump power of the in-loop EDFA, we can adjust the gain, achieving a similar gain profile and peak power for input comb signals with different repetition rates.

As a comparison, we show the change in pulse width, as shown in Fig.\ref{fig:NALM Output}a, where the blue-dashed line shows the results without power optimization and the red-solid line shows the optimized results. By optimizing the pump power, we can maintain the pulse width within 270-280~fs when tuning the tone spacing from 32 to 25~GHz, while it significantly increases to more than 800~fs if the pump power remains the same.
Fig.\ref{fig:NALM Output}b shows the peak to pedestal ratio of the NALM output pulses. Without the power optimization, the ratio varies between 14 to 30~dB across the repetition rate, as illustrated by the blue-dashed line. After power adjustment, the peak to pedestal ratio is improved to a consistent 30~dB, effectively maintaining the pulse quality at different repetition rate, as showed by the red-solid line. Therefore, we can conclude that a maintained pulse and spectral profile can be obtained by adjusting the pump power of the in-loop EDFA provided that the input comb signal has the same pulse shape. However, the tuning capability will be limited by the gain of the bi-directional EDFA, which is determined by the EDF and pump diodes.

\subsection{Parametric comb expansion}
The final output largely depends on the pulse shape and peak power of the pulses into the nonlinear mixer (PM-HNLF2). To achieve this, we employ a zero-dispersion EDFA for pulse amplification and peak power adjustment for different repetition rates. 
This can also be explained using the figure of merit $\varphi_{NL} = \gamma P L$, which describes the maximum phase shift of the pulse peak, where $\gamma$ is the fiber nonlinear coefficient, $P$ is the peak power of the signal and $L$ is the propagation distance. By keeping the pulse shape and peak power the same, one can expect similar pulse shapes and spectral profiles for different tone spacing after the nonlinear mixer.

\begin{figure}[hbt!]
\centering
\includegraphics[scale = 0.24]{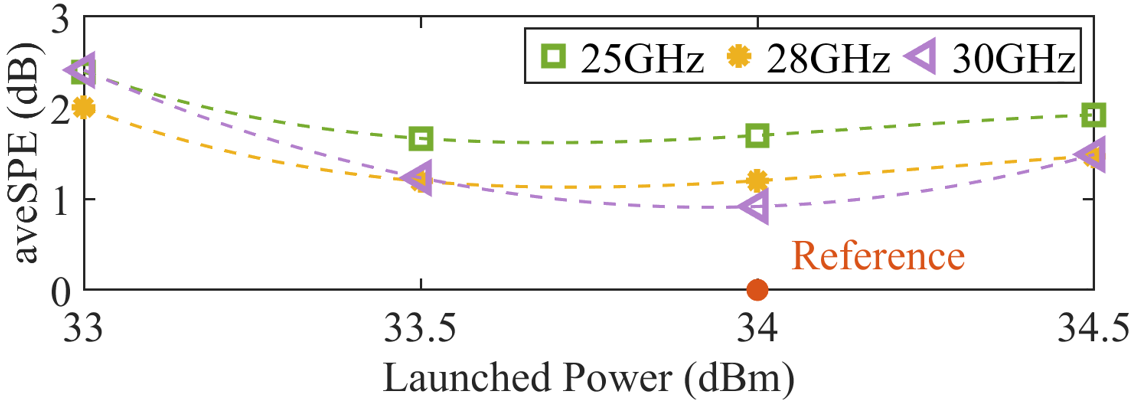}
\caption{Experimental aveSPE for parametric combs of 25, 28 and 30 tone spacing over different launched power, reference point (red circle).}
\label{fig:Parametric comb SPE}
\end{figure}
To show the impact of launch power on spectral profile in the final stage, we show the measured aveSPE at different average launch power into the nonlinear mixer (PM-HNLF~2) using optimized stage 2 outputs, as shown in Fig.\ref{fig:Parametric comb SPE}. 
The reference comb has a tone spacing of 32 GHz, and the launch power is 34 dBm, where the aveSPE is 0~dB. When tuning the tone spacing to 30~GHz, we obtained the optimized spectral profile with an aveSPE of about 1~dB. Further reducing the tone spacing would require slightly reduced launch power at about 33.5~dBm to achieve the lowest aveSPE of 1.3 and 1.8~dB for 28 and 25 GHz tone spacing, respectively. This concludes that pulse energy and shapes need to be maintained to achieve a consistent spectral profile at different tone spacing. 

\begin{figure}[hbt!]
\centering
\includegraphics[scale = 0.166]{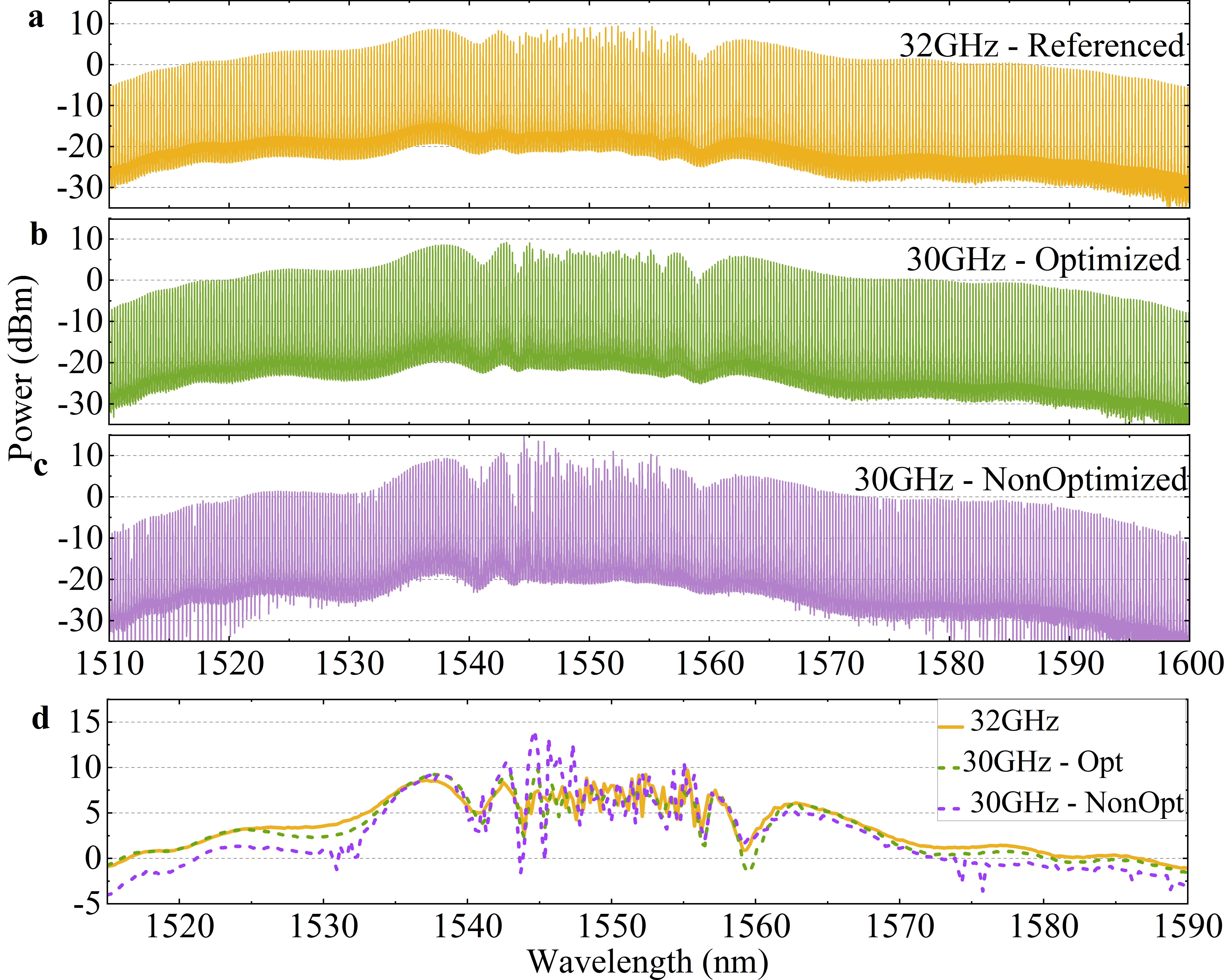}
\caption{Spectra of the nonlinear mixer output \textbf{a.} the reference spectrum at 32~GHz tone spacing. \textbf{b.} optimized 30-GHz spacing. \textbf{c.} 30~GHz tone spacing without power optimization. \textbf{d.} Comparison of the spectral profiles of spectra above.}
\label{fig:Parametric comb spectra}
\end{figure}

Fig.\ref{fig:Parametric comb spectra}a-c shows the measured optical spectra of the 32~GHz and 30~GHz spacing parametric combs with/without power optimization, and the fitted profiles for all three cases are shown in Fig. \ref{fig:Parametric comb spectra}d. By adjusting the parameters in all three stages, we obtain similar spectral profiles at 32~GHz and 30~GHz, which exhibit 10~dB power variation and more than -5~dBm per line over the 1515-1590~nm range. At 32~GHz tone spacing, there are 293 lines and 313 lines when it is tuned to 30-GHz spacing, achieving more than 25~dB OSNR with 0.1~nm noise bandwidth. Fig.\ref{fig:Parametric comb spectra}c shows the measured spectrum by simply tuning the tone spacing to 30~GHz without optimizing operational parameters, where we observe more than a 20~dB power variation of the optical tones within 1510-1600~nm. A strong fluctuation is observed in the center (1540-1560~nm), along with some dips in both the longer and shorter wavelength regions.Fig.\ref{fig:Parametric comb spectra}d shows the fitted spectral profiles of three cases in Fig.\ref{fig:Parametric comb spectra}a-c, allowing us to see more clearly the variation of spectral profiles.
The aveSPE for the non-optimized 30-GHz comb is 2.21~dB, whereas the optimized 30-GHz comb is 0.98~dB. The maxSPE of the non-optimized 30-GHz comb is 10.16~dB, while it is reduced to 3.7~dB for the optimized case.

\subsection{Impact of center wavelength}
\begin{figure}[hbt!]
\centering
\includegraphics[scale = 0.31]{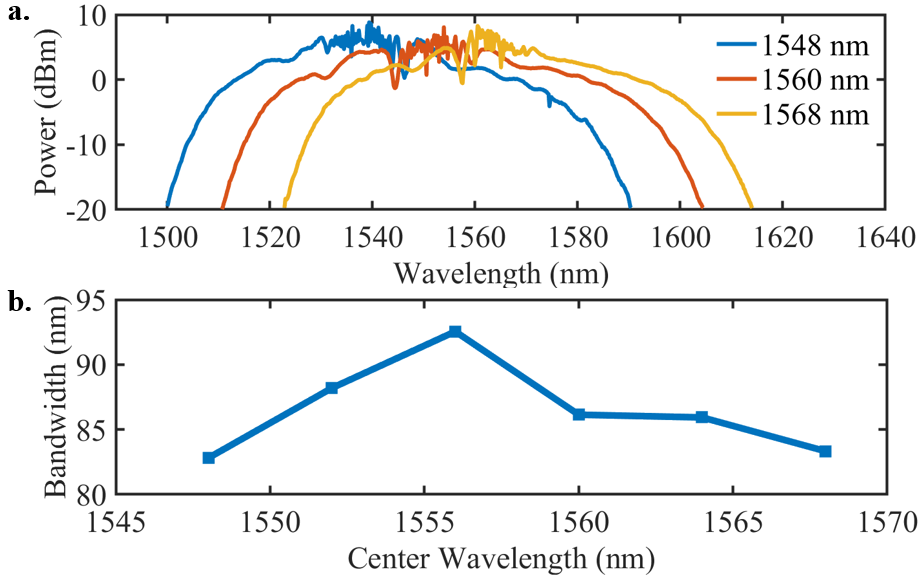}
\caption{Center wavelength tuning of expanded comb: \textbf{a.} Spectral profile with center wavelengths at 1548~nm (blue), at 1560~nm (red), at 1568~nm (yellow); \textbf{b}. 20-dB bandwidth of comb.}
\label{fig:CenTuneSPC}
\end{figure}
In addition to tone spacing tunability, we further study the center wavelength tunability by varying the seed wavelength across 1548-1568 nm. The single-pass EO comb is insensitive to wavelength change. The factors that affect the spectral width and profile in the NALM and the parametric mixer stages include slightly changed fiber dispersion, dispersion slope, and the gain profile of the EDFAs. Here, SPE is no longer a valid metric for evaluating performance due to the varied comb bandwidth. We therefore use a 20-dB bandwidth to indicate the change in spectral width when tuning the wavelength.  
Fig.\ref{fig:CenTuneSPC}a shows the spectral profile of our 25-GHz-spacing comb at different center wavelength, including 1548~nm (blue), 1560~nm (red), and 1568~nm (yellow). The 20-dB bandwidth of the comb over the tuning range is plotted in Fig.\ref{fig:CenTuneSPC}b, which shows a 20-dB bandwidth of more than 83~nm, with the optimum bandwidth being 93~nm when the center wavelength is 1556~nm. This shows that the center wavelength of our cavity-less comb can be continuously tuned without significant change in the spectral profile. The wavelength tuning range is mainly limited by the EDFA operating range. 

\section{Conclusion}
We demonstrate that continuous tuning of tone spacing and center wavelength can be achieved using our cavity-less OFC based on EO-comb and nonlinear expansion. We define spectral profile error to quantitatively study the variations in the spectral profile and show that an aveSPE of less than 2~dB can be achieved by jointly adjusting the RF driving voltage and the amplifier gain in both the NALM and the pulse amplifier. 
The center wavelength can be continuously tuned over 1548-1568~nm, with a 20-dB bandwidth of over 83~nm. 
Such OFC could significantly benefit applications requiring tone spacing and wavelength tunability. 

\begin{backmatter}
    \bmsection{Statement} See supplement 1 for supporting content.
\end{backmatter}
\begin{backmatter}
    \bmsection{Disclosures} The authors declare no conflicts of interest.
\end{backmatter}

\end{document}